\documentclass{Odyssey2026}

 \interspeechcameraready

\title{An Intervention-Based Framework for Shortcut Diagnosis in Spoofing Countermeasures}
\usepackage{amsthm}

\author[affiliation={1}]{Santiago}{Rubio}
\author[affiliation={2}]{Pilar}{Bello}
\author[affiliation={1,2}]{Dayana}{Ribas}
\author[affiliation={1}]{Antonio}{Miguel}
\author[affiliation={1}]{Eduardo}{Lleida}
\author[affiliation={1}]{Alfonso}{Ortega}

\affiliation{ViVoLab, Aragón Institute for Engineering Research (I3A)}{University of Zaragoza}{Spain}
\affiliation{BTS}{Business Telecommunications Services}{Spain}
\email{s.rubio@unizar.es,pbellobardaji@gmail.com,dribas@bts.io, \{amiguel,lleida,ortega\}@unizar.es}

\usepackage{comment}
\usepackage{soul}
\begin{document}

\maketitle

\begin{abstract}

While deepfake audio detection systems achieve high performance in controlled benchmarks, their reliability often diminishes in the wild. Prior work shows that dataset-specific artifacts contribute to this gap. Yet, systematic tools to identify which acoustic properties a model exploits as shortcuts remain limited. We propose an intervention-based diagnostic framework, grounded in a directed graphical model, that formally distinguishes confound-driven shortcut dependencies from legitimate domain shift. We operationalise this through controlled acoustic perturbations targeting non-speech structure, spectral content, and signal energy, complemented by corpus-level distributional analysis. Evaluating XLS-R-300M with RawGAT-ST across ASVspoof challenges datasets, we quantify model sensitivity to specific intervention types. Results reveal that non-speech interventions produce the largest performance shifts, confirming non-speech intervals as a dominant shortcut.

\end{abstract}

\vspace{-6pt}
\section{Introduction}

Detecting synthetic speech has become a critical challenge, and deep learning models have seemingly risen to the occasion, achieving remarkably low error rates on standard benchmarks~\cite{zhu24_asvspoof, xie24_asvspoof}. However, this success is often an illusion. When these highly accurate models are tested in the wild, facing new datasets, different scenarios, or unseen attacks, their performance frequently collapses~\cite{muller22_interspeech, liu2023asvspoof2021, falez24_asvspoof, ASVspoof24}. This persistent generalisation gap undermines confidence across almost all architectures, from traditional classifiers~\cite{rohdin24_asvspoof, combei24_asvspoof, kulkarni24_asvspoof} to the current state-of-the-art based on large self-supervised (SSL) front-ends~\cite{NEURIPS2020_92d1e1eb, Chen_2022, babu2021xlsrselfsupervisedcrosslingualspeech} and advanced back-ends~\cite{Jung2021AASIST, borodin24_asvspoof, dao24_asvspoof, chan24_asvspoof}.

This vulnerability is often driven by ``shortcut learning''~\cite{Geirhos_2020}. A shortcut occurs when a model relies on accidental cues in the dataset that happen to separate real from spoofed audio, rather than learning the actual artifacts of the generation process. If systems depend heavily on these shortcuts, their reported benchmark metrics may dangerously overestimate real-world robustness. Evidence strongly supports this concern: dataset artifacts in anti-spoofing benchmarks were first documented in ASVspoof~2017~\cite{delgado18_odyssey,chettri2020}, and later editions revealed systematic differences in non-speech distributions between bonafide and spoofed audio~\cite{muller21_asvspoof}. The design of ASVspoof~5 explicitly acknowledges this problem by trimming non-speech intervals and normalising energy to mitigate these shortcuts~\cite{wang2024asvspoof5}, a concern echoed in independent analyses~\cite{martindonas24_interspeech}.

Beyond non-speech artifacts, external factors like codec compression and channel effects cause further degradation~\cite{liu2023asvspoof2021, falez24_asvspoof}. While the community has responded with data augmentation, laundering-based simulation~\cite{aliyev24_asvspoof}, guided masking~\cite{truong24_asvspoof}, and synthesis-based diversification~\cite{chen24_asvspoof}, SSL front-ends still retain a consistent advantage under open conditions~\cite{schafer24_asvspoof}. This suggests they naturally absorb part of the robustness burden that augmentation targets in classical systems.

Despite these efforts, the field lacks a formal way to characterise shortcut learning beyond simply documenting individual artifacts~\cite{Geirhos_2020, chettri2020, muller21_asvspoof}. 
Existing work identifies shortcut candidates through empirical 
perturbation studies~\cite{shim23b_interspeech, sahidullah2025shortcut} 
or proposes augmentation-based fixes, but stops short of grounding 
shortcut dependency in the data-generating process. One line of work 
demonstrates that shortcuts can be deliberately constructed or 
suppressed; another quantifies bias through black-box statistical 
frameworks, yet neither provides a formal criterion to determine 
\emph{whether} a performance drop reflects spurious correlation 
exploitation or genuine distribution mismatch, the diagnostic question 
this paper directly addresses.



This paper addresses this diagnostic gap through three contributions.
First, we formalise the data-generating process of spoofing corpora as
a directed graphical model that distinguishes intrinsic synthesis
artifacts~($Z$) from idiosyncratic pipeline choices~($C_d$) and
exogenous channel effects~($C_i$), and derive a definition of
confounded shortcut dependency grounded in the conditional independence
structure of the graph.  Second, we operationalise this framework
through a battery of controlled acoustic perturbations, targeting
non-speech structure, spectral content, and signal energy, designed to
selectively modify candidate shortcut features while preserving the
intrinsic generative content, and measure sensitivity through the relative degradation in detection cost under each perturbation.  Third, we apply the complete framework to a hybrid architecture pairing
XLS-R-300M~\cite{babu2021xlsrselfsupervisedcrosslingualspeech} with
RawGAT-ST~\cite{Jung2021AASIST,tak21_asvspoof} across ASVspoof 2019
LA~\cite{ASVspoof19}, ASVspoof 2021 LA~\cite{liu2023asvspoof2021}, and
ASVspoof~5~\cite{ASVspoof24}, evaluating five training configurations that 
span frozen and fine-tuned SSL front-ends, with and without data augmentation, 
to isolate the effect of front-end adaptation on shortcut sensitivity.

\begin{figure}[t]
\centering
\begin{tikzpicture}[
    >=Stealth,
    obs/.style={
        circle, draw=black, thick, fill=black!10,
        minimum size=9mm, inner sep=0pt, font=\small
    },
    lat/.style={
        circle, draw=black, thick, dashed, fill=white,
        minimum size=9mm, inner sep=0pt, font=\small
    },
    lat_estad/.style={
        circle, draw=black, thick, fill=white,
        minimum size=9mm, inner sep=0pt, font=\small
    },
    partobs/.style={
        circle, draw=black, thick, fill=black!10,
        minimum size=9mm, inner sep=0pt, font=\small,
        path picture={
            \fill[white] 
                (path picture bounding box.west)
                rectangle 
                (path picture bounding box.center
                 |- path picture bounding box.north);
            \fill[white] 
                (path picture bounding box.center
                 |- path picture bounding box.south)
                rectangle 
                (path picture bounding box.east);
        }
    },
    plate/.style={
        rectangle, draw=black!70, thin,
        rounded corners=1pt, fill=none,
        inner sep=10pt
    },
    edge/.style={->, thick, draw=black},
    inter/.style={->, thick, dashed, draw=orange!70!black},
]

\node[lat_estad] (M) at ( 0.0,  4.8) {$M$};
\node[obs] (S) at ( 2.4,  4.8) {$S$};

\node[lat_estad] (Cd) at (-2.2,  2.8) {$C_d$};
\node[lat_estad]     (Z)  at ( 0.0,  2.8) {$Z$};
\node[lat_estad] (Ci) at ( 2.2,  2.8) {$C_i$};

\node[plate, fit=(Cd), minimum width=1.5cm, minimum height=1.5cm]
    (plateCd) {};
\node[font=\normalsize, anchor=south east, inner sep=1pt]
    at (plateCd.south east) {$D$};

\node[plate, fit=(Ci), minimum width=1.5cm, minimum height=1.5cm]
    (plateCi) {};
\node[font=\normalsize, anchor=south east, inner sep=1pt]
    at (plateCi.south east) {$I$};

\node[obs] (x) at ( 0.0,  1.1) {$\mathbf{x}$};


\node[lat_estad] (Zhat) at ( 0.0, -0.2) {$\hat{Z}$};

\node[obs] (Shat) at ( 0.0, -1.5) {$\hat{S}$};

\node[obs, fill=white, draw=orange!80!black, thick]
    (H) at (-2.2, 4.8) {$H$};

\draw[edge] (M) -- (S);
\draw[edge] (M) to[out=255, in=90] (Z);
\draw[edge] (M) to[out=220, in=80] (Cd);

\draw[edge] (Z)  to[out=270, in=90]  (x);
\draw[edge] (Cd) to[out=300, in=150] (x);
\draw[edge] (Ci) to[out=240, in=30]  (x);

\draw[edge] (x) to[out=270, in=90]  (Zhat);
\draw[edge] (Zhat) to[out=270, in=90]  (Shat);

\draw[inter] (H) to[out=270, in=90] (plateCd.north);
\draw[inter] (H) to[out=310, in=90] (plateCi.north);

\end{tikzpicture}
\vspace{-6.5pt}
\caption{
    Directed graphical model of the data-generating process and
    diagnostic framework~\cite{bishop2006pattern,pearl2009causality}.
    Shaded nodes: observed ($\mathbf{x}$, $S$, $\hat{S}$);
    Solid nodes: latent ($M$, $Z$, $C_d$, $C_i$, $\hat{Z}$).
    Plates $D$ and $I$ index the sets of idiosyncratic and exogenous
    factors.  Orange dashed arrows denote controlled
    interventions~$H$
}
\vspace{-6.5pt}
\label{fig:pgm_full}
\end{figure}
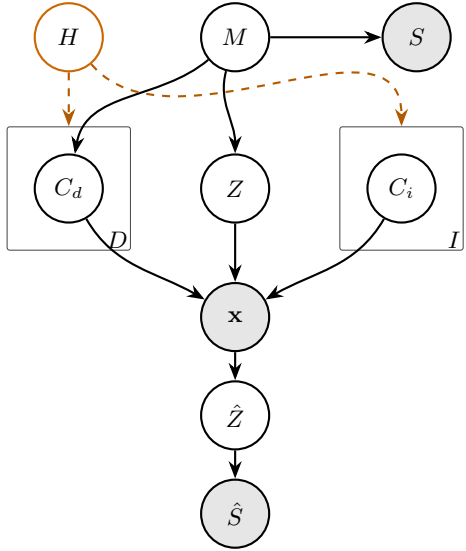

\section{Problem Formulation}
\label{sec:problem_formulation}
\vspace{-2pt}
\subsection{Task Definition and Notation}
\label{sec:task}
\vspace{-2pt}
Let $\mathbf{x} \in \mathbb{R}^{T}$ denote a raw speech waveform of $T$
samples and let $S \in \{0,1\}$ indicate the true nature of the utterance,
where $S{=}1$ denotes bonafide and $S{=}0$ spoofed. A hybrid spoofing
countermeasure learns a mapping $f_\theta \colon \mathbb{R}^{T} \to [0,1]$,
decomposed as $ f_\theta =  h_\psi \circ g_\phi ,$
where $g_\phi$ is a self-supervised front-end that produces a sequence of
frame-level representations $\mathbf{R} = g_\phi(\mathbf{x})$, and $h_\psi$ is
a downstream classifier that maps $\mathbf{R}$ to a scalar score
$s = h_\psi(\mathbf{R}) \in \mathbb{R}$, yielding a binary prediction
$\hat{S} = \mathbf{1}[s > \tau]$ for threshold~$\tau$.  We denote by
$\hat{Z}$ the internal utterance-level representations learned by the model; whether $\phi$
is held frozen or updated during training constitutes a variable of the
analysis. The parameter set $\theta = \{\phi, \psi\}$ is optimised over a training corpus $\mathcal{D}_\mathrm{train} = \{(\mathbf{x}_i, S_i)\}_{i=1}^n$.

\vspace{-8pt}
\subsection{Structure of the Data Generation Process}
\label{sec:causal_dgp}

We frame the data generation and annotation as a graphical model, represented by the directed acyclic graph in Fig.~\ref{fig:pgm_full}. The generative mechanism $M$ sets the true nature $S$ of 
the utterance: if $M$ is a human vocal tract, then bonafide; if $M$ is a TTS or VC pipeline, then spoof. The corpus-assigned label $S_i \in \{0, 1\}$ is a protocol-dependent realisation that reflects the labelling decisions of a specific dataset rather than a direct measurement of the underlying generative process. The mechanism $M$ leaves two structurally distinct traces in the observed waveform:
\begin{itemize}
    \item $Z$: \textbf{Intrinsic Artifacts.} Fundamental ``fingerprint'' inherent to the synthesis process regardless of the specific implementation (e.g., vocoder phase anomalies). These are the robust, generalizable features that models should ideally learn.
    \item $C_d$: \textbf{Idiosyncratic Artifacts.} Extrinsic acoustic characteristics dependent on the specific generation pipeline (e.g., unnatural non-speech distributions or peak normalization). While generated by $M$, they are extrinsic algorithmic choices rather than fundamental properties of spoofed speech.
\end{itemize}
Additionally, the observed waveform $\mathbf{x}$ can be modulated by class-independent exogenous factors $C_i$ (e.g., codecs or tx channel) which are causally disconnected from $M$. Crucially, the protocol design of any specific training corpus $\mathcal{D}_{\mathrm{train}}$ inevitably introduces a \textit{spurious correlation} between the idiosyncratic artifacts $C_d$ and the label $S$. Because datasets contain a limited diversity of generative pipelines, certain post-processing artifacts become strongly correlated with the spoofed class.

\subsection{Confounded Shortcut Dependencies}
\label{sec:confounded_shortcuts}

Cross-condition degradation stems from two distinct failure modes: \textbf{Domain shift} arises when the distribution of one or more exogenous variables differs between training and evaluation. In its most common form, a change in $P(C_i)$ alters the marginal $P(\mathbf{x})$. This naturally degrades the performance of almost any model, but it does not mean the model learned the wrong features.

\textbf{Shortcut learning}, on the other hand, occurs when the model re encode spurious statistical regularities of the training corpus~\cite{Geirhos_2020}. We formalise this with
respect to the DAG in Fig.~\ref{fig:pgm_full}.  Let $\hat{Z}$ denote
the internal representation learned by the model.  The ideal
countermeasure satisfies the \emph{causal sufficiency
condition}~\cite{pearl2009causality}:
{ 
\setlength{\abovedisplayskip}{5pt}
\setlength{\belowdisplayskip}{5pt}
\begin{equation}\label{eq:sufficiency}
    \hat Z \indep (C_d,C_i)~|~Z
\end{equation}
}%
that is, the model's representation depends on the observed
waveform~$\mathbf{x}$ only through the intrinsic artifacts~$Z$,
discarding both idiosyncratic pipeline choices~$C_d$ and exogenous
channel effects~$C_i$.

We say that the model exhibits a \emph{confounded shortcut dependency} on~$C_d$ if the following conditions hold jointly:
\begin{enumerate}

    \item[(i)] \textbf{Confound-driven association.}  Under the
    training distribution, $C_d$ is statistically associated with the
    label~$S$:
    { 
    \setlength{\abovedisplayskip}{5pt}
    \setlength{\belowdisplayskip}{5pt}
    \begin{equation}\label{eq:spurious}
        C_d \not\indep S \mid \mathcal{D}_{\mathrm{train}}
    \end{equation}
    }%
    Crucially, this is not a true cause-and-effect relationship. The artifact $C_d$ does not define whether an audio is a deepfake. Instead, this accidental correlation appears only because  $\mathcal{D}_{\mathrm{train}}$ contains a limited diversity
    of mechanisms~$M$, creating a confound-driven statistical pathway
    from~$C_d$ to~$S$.


    \item[(ii)] \textbf{Representational leakage.}  The learned
    representation violates the sufficiency condition
    (Eq.~\ref{eq:sufficiency}) by encoding~$C_d$:
    { 
    \setlength{\abovedisplayskip}{5pt}
    \setlength{\belowdisplayskip}{5pt}    
    \begin{equation}\label{eq:leakage}
        \hat{Z} \not\indep C_d \mid Z
    \end{equation}
    }
    Rather than relying on the path
    $Z \to \mathbf{x} \to \hat{Z} \to \hat{S}$, the model exploits
    $C_d \to \mathbf{x} \to \hat{Z} \to \hat{S}$ --- a path that is
    predictive under $\mathcal{D}_{\mathrm{train}}$ but unstable under
    distribution shift, precisely because the association
    $C_d$--$S$ is conditional on the training protocol rather than on
    the generative structure. This shortcut inflates benchmark scores but fails in the wild where this correlation vanishes.
    
\end{enumerate}
As a result, when $P_\mathrm{eval}(C_d \mid S) \neq P_\mathrm{train}(C_d \mid S)$, the confound-driven association ceases to be predictive and performance degrades. This fragility is the observable symptom, not the definition, of the underlying failure. Unlike domain shift, which degrades any model and is typically addressed via data augmentation, shortcut degradation is a selective failure. A model relying on $C_d$ will misclassify novel attacks not because the acoustic environment has changed, but because its decision boundary was never grounded in the generative structure of the task~\cite{wang2024asvspoof5}. Diagnosing these shortcut dependencies requires controlled interventions that alter an estimation of $C_d$ while preserving $Z$.


\subsection{Diagnostic Strategy: From Interventions to Shortcut Evidence}
\label{sec:diagnostic_strategy}
 
Our framework defines a confounded shortcut dependency using two conditions: spurious dataset correlation (Eq.~\ref{eq:spurious}) and representational leakage (Eq.~\ref{eq:leakage}). While the first condition can be verified through corpus-level statistical analysis, the second requires empirical proof that the learned representation $\hat{Z}$
encodes~$C_d$. This section describes how controlled acoustic
perturbations provide such evidence.
 
The ideal test would directly modify the idiosyncratic artefact~$C_d$
while holding $Z$ fixed and observing whether the model prediction
changes. In practice, $C_d$ is latent and cannot be
manipulated in isolation.What we can manipulate is the observed
waveform~$\mathbf{x}$.  A key property of the framework makes this
tractable: the intrinsic artifacts $Z$ are imprinted in the waveform at
generation time by the synthesis mechanism $M$, and no post-hoc
perturbation applied to $\mathbf{x}$ can alter them.  Adding non-speech fragments,
injecting noise, or filtering frequency bands cannot reverse or modify the
acoustic fingerprint left by a vocoder or a TTS acoustic model.
Perturbations can only modify $C_d$ (e.g., non-speech structure),
$C_i$ (e.g., channel characteristics), or \emph{mask} the observability
of $Z$ (e.g., by destroying the spectral region where $Z$ is most
evident), but they cannot change $Z$ itself.
 
This distinction yields a clear inferential logic. This leads to a clear diagnostic logic: if we perturb a candidate artifact $C_d$ without masking $Z$, and the model's performance drops, the degradation reflects the model's reliance on that shortcut. Non-speech perturbations are the perfect tool for this. They operate on non-speech regions where intrinsic generative traces are typically minimal or easily overwritten by trivial post-processing. By modifying $C_d$ without altering or masking $Z$, they provide an exceptionally clean diagnostic for shortcut dependence. In contrast, spectral and some energy perturbations are more ambiguous; a performance drop could mean either shortcut exploitation or the accidental masking of legitimate $Z$ features.

Our diagnostic protocol proceeds as follows: for each candidate shortcut identified in the corpus (Section~\ref{sec:corpus_analysis}), we design a targeted perturbation or intervention, apply it to the evaluation data, and measure the resulting performance shift using relative DCF degradation ($\delta_{m,p}$) (Section~\ref{sec:metrics}). Comparing these shifts across non-speech, spectral, and energy perturbations yields a \emph{sensitivity profile} that precisely characterises which acoustic properties the model is actively exploiting.

\section{Experimental Setup}
\subsection{Architectural Configuration}
\label{sec:architecture}

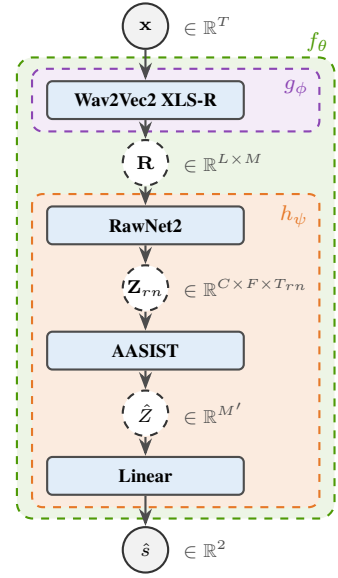
\begin{figure}[t]
    \centering
    \begin{tikzpicture}[
        >=Stealth,
        node distance=0.25cm,
        obs/.style={circle, draw=black!80, thick, fill=black!5, minimum size=6mm, inner sep=0pt, font=\scriptsize},
        lat/.style={circle, draw=black!80, thick, dashed, fill=white, minimum size=6mm, inner sep=0pt, font=\scriptsize},
        blk/.style={rectangle, draw=black!70, thick, rounded corners=2pt, minimum height=5mm, minimum width=26mm, align=center, font=\scriptsize\bfseries, fill=myblue!40},
        edge/.style={->, thick, draw=black!70},
        dim/.style={font=\scriptsize, text=black!70}
    ]

        \node[obs] (x) {$\mathbf{x}$};
        
        \node[blk, below=0.4cm of x] (xlsr) {Wav2Vec2 XLS-R};
        \node[lat, below=of xlsr] (R) {$\mathbf{R}$};
        \node[blk, below=of R] (rawnet) {RawNet2};
        \node[lat, below=of rawnet] (Zrn) {$\mathbf{Z}_{rn}$};
        \node[blk, below=of Zrn] (aasist) {AASIST};
        \node[lat, below=of aasist] (Zhat) {$\hat{Z}$};
        \node[blk, below=of Zhat] (linear) {Linear};
        
        \node[obs, below=0.4cm of linear] (shat) {$\hat{s}$};

        \node[dim, right=2pt of x, anchor=west] {$\in \mathbb{R}^{T}$};
        \node[dim, right=2pt of R, anchor=west] {$\in \mathbb{R}^{L \times M}$};
        \node[dim, right=2pt of Zrn, anchor=west] {$\in \mathbb{R}^{C \times F \times T_{rn}}$};
        \node[dim, right=2pt of Zhat, anchor=west] {$\in \mathbb{R}^{M^{\prime}}$};
        \node[dim, right=2pt of shat, anchor=west] {$\in \mathbb{R}^{2}$};

        \draw[edge] (x) -- (xlsr);
        \draw[edge] (xlsr) -- (R);
        \draw[edge] (R) -- (rawnet);
        \draw[edge] (rawnet) -- (Zrn);
        \draw[edge] (Zrn) -- (aasist);
        \draw[edge] (aasist) -- (Zhat);
        \draw[edge] (Zhat) -- (linear);
        \draw[edge] (linear) -- (shat);

        \begin{scope}[on background layer]
            \coordinate (LeftF)  at (-1.7, 0); 
            \coordinate (RightF) at (2.5, 0);

            \coordinate (LeftGH)  at (-1.5, 0);
            \coordinate (RightGH) at (2.3, 0);

            \coordinate (f_nw) at ($(LeftF |- xlsr.north) + (0, 0.3)$);
            \coordinate (f_se) at ($(RightF |- linear.south) + (0, -0.3)$);
            \filldraw[draw=mygreen, dashed, thick, fill=mygreen!10, rounded corners=6pt]
                (f_nw) rectangle (f_se);
            \node[anchor=south east, text=mygreen, font=\footnotesize\bfseries, inner sep=2pt] at (RightF |- f_nw) {$f_\theta$};

            \coordinate (g_nw) at ($(LeftGH |- xlsr.north) + (0, 0.15)$);
            \coordinate (g_se) at ($(RightGH |- xlsr.south) + (0, -0.15)$);
            \filldraw[draw=mypurple, dashed, thick, fill=mypurple!10, rounded corners=4pt]
                (g_nw) rectangle (g_se);
            \node[anchor=north east, text=mypurple, font=\footnotesize\bfseries, inner sep=4pt] at (RightGH |- g_nw) {$g_\phi$};

            \coordinate (h_nw) at ($(LeftGH |- rawnet.north) + (0, 0.15)$);
            \coordinate (h_se) at ($(RightGH |- linear.south) + (0, -0.15)$);
            \filldraw[draw=myorange, dashed, thick, fill=myorange!15, rounded corners=4pt]
                (h_nw) rectangle (h_se);
            \node[anchor=north east, text=myorange, font=\footnotesize\bfseries, inner sep=4pt] at (RightGH |- h_nw) {$h_\psi$};
        \end{scope}

    \end{tikzpicture}
    \vspace{-7pt}
    \caption{Detailed architecture of the proposed hybrid SSL-based spoofing detection model. Intermediate latent representations ($\mathbf{R}$, $\mathbf{Z}_{rn}$, $\hat{Z}$) and dimensionalities are mapped along the data flow from input $\mathbf{x}$ to the final score $\hat{s}$.}
    \label{fig:arquitectura}
    \vspace{-18pt}
\end{figure}

The hybrid countermeasure follows the
$f_\theta = h_\psi \circ g_\phi$ formulation (Section~\ref{sec:task},
Fig.~\ref{fig:arquitectura}).  The front-end $g_\phi$ is
XLS-R-300M~\cite{babu2021xlsrselfsupervisedcrosslingualspeech}, held frozen or fine-tuned.  The classifier $h_\psi$ follows the RawGAT-ST
architecture~\cite{tak21_asvspoof}, producing a time-independent
embedding $\hat{Z} \in \mathbb{R}^{160}$ mapped to two class
logits.

\vspace{-12pt}
\paragraph*{\textbf{Training and Evaluation Corpora}}The primary training corpus is ASVspoof 2019
LA\footnote{For brevity, all subsequent references to ASVspoof 2019 and 2021 refer exclusively to their Logical Access (LA) partitions.}~\cite{ASVspoof19}, comprising spoofed utterances from 6~TTS and voice conversion systems. The evaluation partition covers 13~additional unseen synthesis algorithms.  A subset of models is additionally trained on
a joint corpus combining ASVspoof 2019~LA and ASVspoof~5
training data~\cite{wang2024asvspoof5}; the ASVspoof~5 training
partition contains 8~TTS/VC systems, drawn from a broader set of
32~algorithms spanning the full corpus. All models are evaluated under three conditions of increasing domain
distance: (i)~\textbf{ASVspoof 2019 LA eval}~\cite{ASVspoof19}, in-domain, unseen
algorithms but matched recording conditions;
(ii)~\textbf{ASVspoof 2021 LA eval}~\cite{liu2023asvspoof2021},
same TTS systems transmitted through telephone and VoIP codecs,
introducing channel variability absent from training; and
(iii)~\textbf{ASVspoof 5 eval}~\cite{ASVspoof24}, fully disjoint synthesis
algorithms under crowdsourced and adversarially perturbed conditions.

\vspace{-12pt}
\paragraph*{\textbf{Training configuration.}}
All input waveforms are peak-normalised and adjusted to 64{,}600
samples (${\approx}\,4$\,s at 16\,kHz).  Models are trained for
30~epochs with batch size~24, Adam optimiser (lr\,$= 10^{-6}$, weight
decay\,$= 10^{-4}$), and weighted cross-entropy with
$(w_\mathrm{sp}, w_\mathrm{bona}) = (0.1,\,0.9)$.  Two augmentation
regimes are compared. The first applies \textit{RawBoost}~(RB)~\cite{tak2021rawboost}, which
introduces convolutive and impulsive additive noise on the raw
waveform, with standard truncation/tiling for length normalisation. The second is a \textit{custom pipeline}~(DA), inspired by strategies
from recent ASVspoof~5 submissions~\cite{schafer24_asvspoof}, that applies a stochastic chain of augmentations: non-speech trimming, RIR
convolution~\cite{ko2017study}, background noise addition at
8--20\,dB SNR~\cite{app12189000}, RawBoost, codec simulation matching
ASVspoof 2021~LA conditions~\cite{liu2023asvspoof2021}, and time
masking.  Each stage is applied independently with a fixed probability,
and 20\% of samples pass through unchanged to preserve clean examples
in the training distribution. Length normalisation in DA uses reflection padding with random circular shift instead of tiling.

\subsection{Corpus-Level Distributional Analysis}
\label{sec:corpus_analysis}

\begin{figure*}[t]
    \centering
    \includegraphics[width=0.9\linewidth]{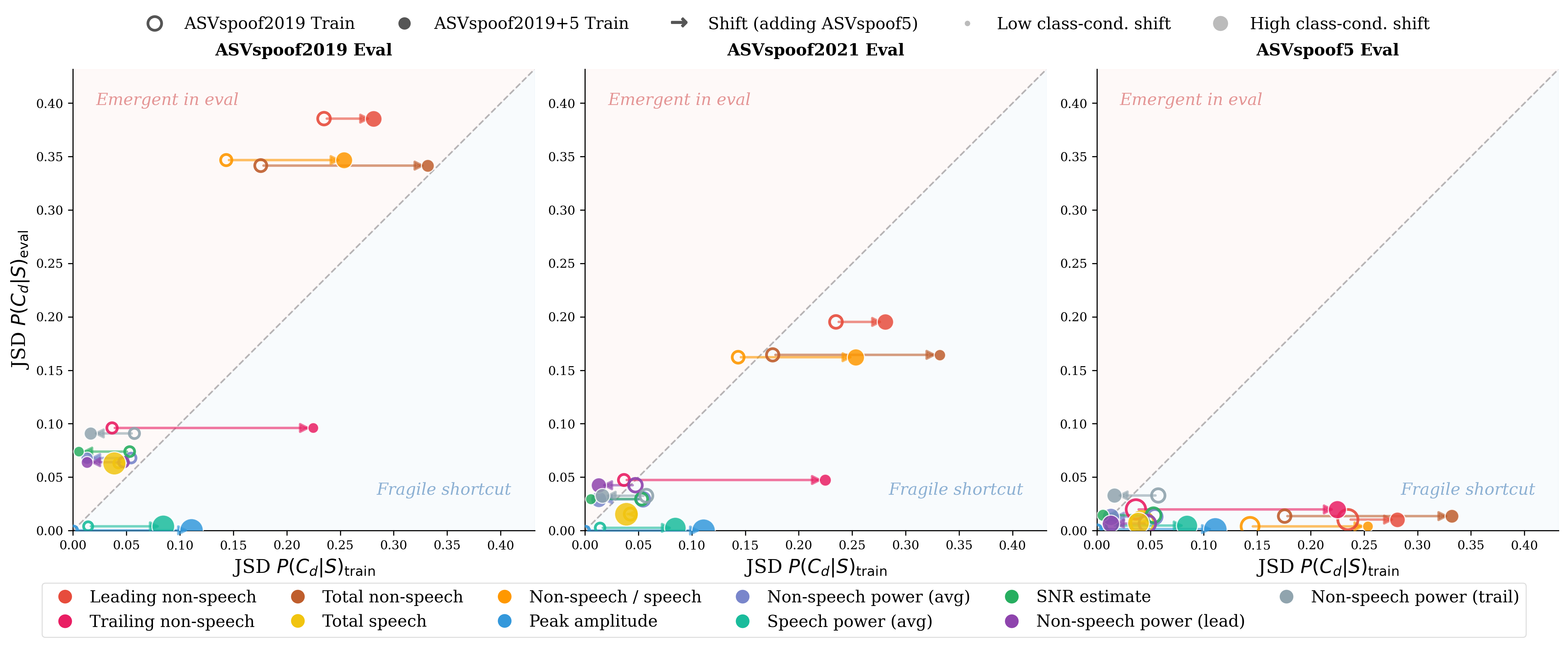}
    \vspace{-8pt}
    \caption{Class separability across conditions: JSD between 
    bonafide and spoofed distributions on the training partition 
    ($x$-axis) vs.\ each evaluation set ($y$-axis). Hollow/filled 
    markers: trained on ASV19 only / ASV19+ASV5; arrows: shift from 
    adding AS5 data. Points in the lower-right are separable in 
    training but not in evaluation (fragile shortcut).}
    \label{fig:scatter_jsd}
    \vspace{-12pt}
\end{figure*}

We extract eleven acoustic descriptors per utterance using Silero VAD~\cite{silero_vad}. To identify candidate shortcuts, Figure~\ref{fig:scatter_jsd} plots the Jensen--Shannon divergence (JSD)~\cite{61115} between bonafide and spoofed acoustic descriptors distributions for each descriptor. The $x$-axis shows separability in the training set, while the $y$-axis shows separability in the evaluation sets (high JSD denotes strong class separability). Additionally, the circle size encodes the distribution shift between train and evaluation sets: large circles indicate unstable properties across domains, whereas small circles denote stable descriptors.

In the ASVspoof 2019 panel, non-speech-related descriptors cluster in the upper-right, meaning they highly separate classes in both training and evaluation. However, they shift toward the lower-right in 2021, and nearly vanish in ASVspoof 5. This drop aligns perfectly with the ASVspoof 5 protocol, which explicitly trims non-speech regions and normalises energy in its evaluation data~\cite{wang2024asvspoof5}. This pattern confirms non-speech intervals as a fragile shortcut: highly exploitable during training, but progressively useless as evaluation conditions change.

The arrows reveal a counterintuitive effect: adding ASVspoof~5 training data increases the training-side bias for both non-speech and certain energy descriptors. This occurs because this training partition preserves the non-speech and energy asymmetries that its evaluation protocol suppresses~\cite{wang2024asvspoof5}. Therefore, blindly expanding a corpus simply amplifies the confound if the new data carries the same biases. In contrast, signal quality descriptors remain stably near the origin, confirming they do not act as shortcuts regardless of the training corpus.

\begin{figure}[!h]
    \centering
    \includegraphics[width=0.85\linewidth]{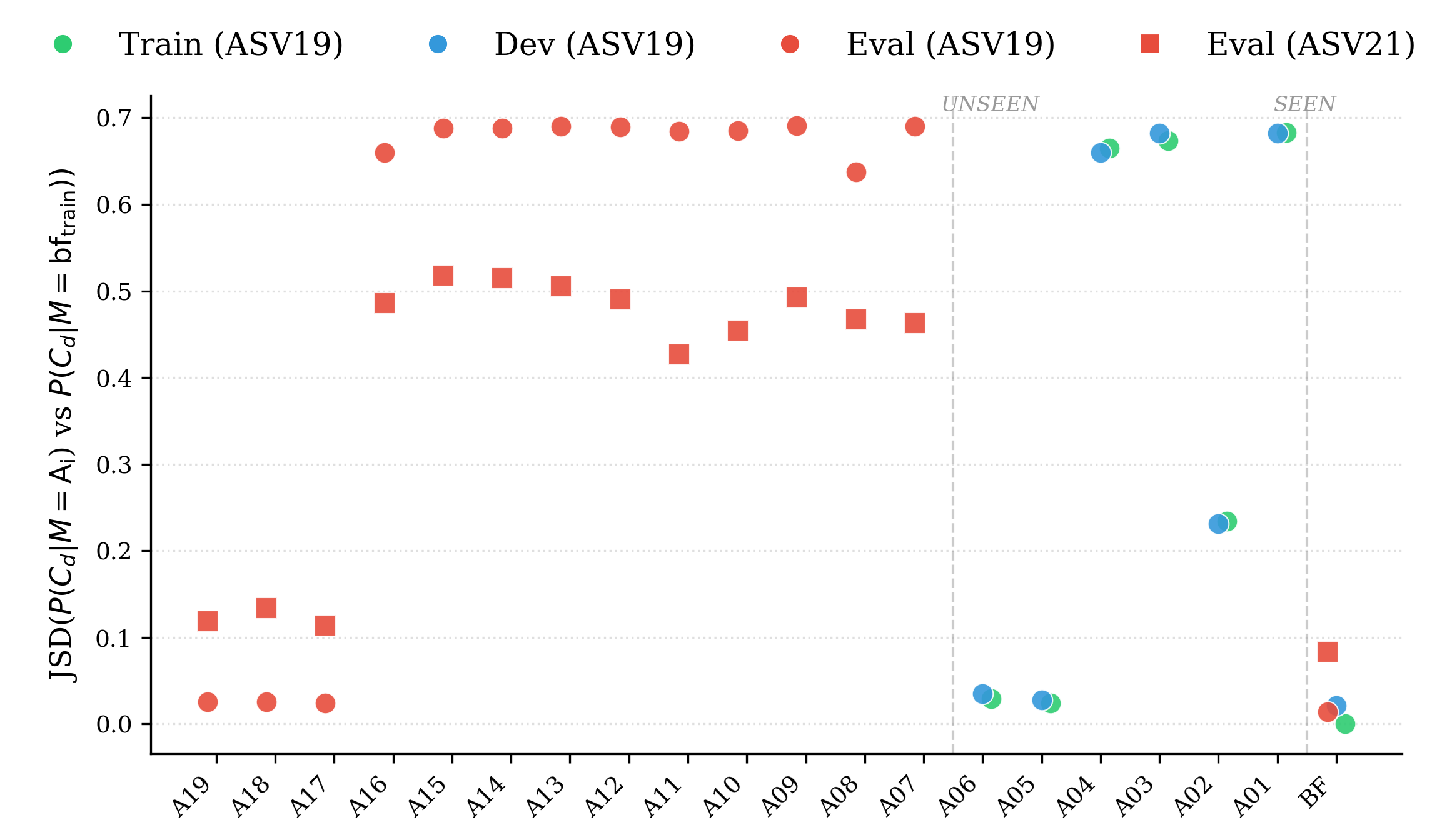}
    \vspace{-7pt}
    \caption{Per-mechanism JSD for leading\_non-speech\_dur in ASV19 and ASV21, computed between the training bonafide distribution ASV19 and each attack or evaluation bonafide set.}
    \label{fig:per_attack_jsd}
    \vspace{-8pt}
\end{figure}

To understand this further, Figure~\ref{fig:per_attack_jsd} breaks down the JSD for leading non-speech regions, one of the strongest shortcuts, by specific attack algorithms across the 2019 and 2021 datasets. The high variance across different attacks confirms that the shortcut strength depends heavily on the specific synthesis mechanism ($M$). Notably, the extreme separability observed for several attacks indicates they could be trivially classified by this property alone. Furthermore, the bonafide distribution in ASVspoof 2021 shifts drastically because its \texttt{only\_speech} annotation~\cite{liu2023asvspoof2021} artificially strips leading non-speech regions. This reduction in class separability occurs independently of the spoofed class, proving that the shortcut is tied to the dataset's protocol rather than the generative process itself.

\subsection{Intervention Framework}
\label{sec:intervention_design}
Following the strategy in Section~\ref{sec:diagnostic_strategy}, we design controlled interventions targeting three acoustic categories: non-speech structure, spectral content, and signal energy. Table~\ref{tab:perturbations} summarises the set used in the reported experiments.

Non-speech manipulations constitute the largest group, reflecting the strong distributional asymmetry documented in Section~\ref{sec:corpus_analysis} .We inject 4.0\,s of either zero-padding or AWGN (Additive White Gaussian Noise) at the beginning (lead) or end (trail) of the audio, substantially exceeding the non-speech durations observed in the training corpus. Spectral and energy interventions probe dependence on frequency content and amplitude/noise robustness, respectively. For peak normalisation, the target is drawn from a narrow distribution centred at the nominal value (0.65 or 0.45) to avoid introducing a perfectly deterministic artificial cue. Intervention-based diagnosis is conducted only on ASVspoof 2019 and 2021.

\begin{table}[!ht]
\vspace{-6pt}
\centering
\caption{Controlled intervention set for shortcut probing, grouped by
targeted acoustic property.}
\vspace{-7pt}
\label{tab:perturbations}
\setlength{\tabcolsep}{4pt}
\small
\begin{tabular}{@{}lll@{}}
\toprule
\textbf{Category} & \textbf{Perturbation} & \textbf{Parameters} \\
\midrule
\multirow{2}{*}{Non-Speech}
  & Zero padding   & lead / trail, 4.0\,s \\
  & AWGN padding   & lead / trail, 4.0\,s \\
\midrule
\multirow{2}{*}{Spectral}
  & Band-cut filter & 0--2 / 2--5 / 5--8\,kHz \\
  & Downsample      & 8\,kHz \\
\midrule
\multirow{2}{*}{Energy}
  & Additive AWGN    & 20 / 10 / 5\,dB SNR \\
  & Peak normalisation & $\mu = 0.65$ / $\mu = 0.45$ \\
\bottomrule
\end{tabular}
\end{table}

\begin{table*}[t]
\centering
\resizebox{\linewidth}{!}{
\begin{tabular}{l l c c | c c c c | c c c c c | c c c c}
\toprule
\multirow{2}{*}{} & \multirow{2}{*}{\textbf{Model}} & \multirow{2}{*}{\textbf{Inter.}} & \multirow{2}{*}{\textbf{Base DCF}} & \multicolumn{4}{c|}{\textbf{Spectral}} & \multicolumn{5}{c|}{\textbf{Noise/Amp}} & \multicolumn{4}{c}{\textbf{Non-Speech}} \\
\cmidrule(lr){5-8} \cmidrule(lr){9-13} \cmidrule(lr){14-17}
 &  &  &  & \textit{S1} & \textit{S2} & \textit{S3} & \textit{S4} & \textit{N1} & \textit{N2} & \textit{N3} & \textit{N4} & \textit{N5} & \textit{T1} & \textit{T2} & \textit{T3} & \textit{T4} \\
\midrule
\multirow{10}{*}{\rotatebox[origin=c]{90}{\textbf{ASVspoof 2019 eval}}}  & \multirow{2}{*}{RB-19*} & \textit{Spoof} & \multirow{2}{*}{0.109} & -0.34 & +0.08 & -0.17 & -0.23 & +0.56 & +0.59 & +0.54 & +0.02 & +0.04 & +1.43 & +1.40 & +0.81 & +0.39 \\
 &  & \textit{Both} &  & \textbf{+15.96} & +0.30 & \textbf{+1.39} & \textbf{+3.06} & +0.29 & +1.90 & +4.02 & +0.06 & +0.30 & +0.82 & +0.77 & +0.34 & -0.07 \\
\cmidrule(lr){2-17}
 & \multirow{2}{*}{RB-19} & \textit{Spoof} & \multirow{2}{*}{0.004} & +14.25 & -0.14 & -0.26 & -0.28 & +0.80 & +2.85 & +5.20 & +0.13 & +0.22 & \textbf{+63.73} & +43.73 & \textbf{+65.95} & +44.25 \\
 &  & \textit{Both} &  & \textbf{+32.99} & -0.14 & +0.07 & +0.25 & +0.67 & +2.65 & +5.07 & +0.13 & +0.22 & \textbf{+63.07} & +43.08 & \textbf{+65.29} & +43.60 \\
\cmidrule(lr){2-17}
 & \multirow{2}{*}{DA-19} & \textit{Spoof} & \multirow{2}{*}{0.203} & -0.35 & +0.06 & +0.02 & +0.03 & +0.04 & +0.04 & -0.05 & +0.07 & +0.13 & +0.49 & +0.39 & +0.42 & +0.36 \\
 &  & \textit{Both} &  & \textbf{+8.03} & +0.07 & +0.05 & +0.05 & +0.05 & +0.48 & \textbf{+1.13} & -0.12 & -0.17 & -0.11 & -0.19 & -0.10 & -0.15 \\
\cmidrule(lr){2-17}
 & \multirow{2}{*}{RB-24} & \textit{Spoof} & \multirow{2}{*}{0.006} & +6.96 & +1.26 & +0.14 & +0.05 & +1.14 & +1.26 & +1.43 & +0.55 & +1.28 & \textbf{+15.60} & +4.03 & \textbf{+18.83} & +2.84 \\
 &  & \textit{Both} &  & \textbf{+71.43} & +4.66 & +0.58 & +2.84 & +1.01 & \textbf{+9.28} & \textbf{+49.66} & +0.38 & +1.11 & +14.94 & +4.94 & +18.39 & +13.95 \\
\cmidrule(lr){2-17}
 & \multirow{2}{*}{DA-24} & \textit{Spoof} & \multirow{2}{*}{0.261} & -0.06 & +0.11 & +0.06 & +0.02 & -0.00 & -0.11 & -0.23 & +0.07 & +0.12 & -0.22 & -0.22 & -0.27 & -0.23 \\
 &  & \textit{Both} &  & +3.78 & +0.16 & +0.05 & +0.29 & +1.56 & \textbf{+3.64} & \textbf{+5.33} & -0.20 & -0.26 & +0.98 & +1.00 & \textbf{+2.47} & \textbf{+2.20} \\
\midrule
\multirow{10}{*}{\rotatebox[origin=c]{90}{\textbf{ASVspoof 2021 eval}}}  & \multirow{2}{*}{RB-19*} & \textit{Spoof} & \multirow{2}{*}{0.736} & -0.03 & +0.01 & -0.01 & -0.02 & +0.07 & +0.09 & +0.09 & -0.00 & -0.00 & +0.16 & +0.17 & +0.09 & +0.04 \\
 &  & \textit{Both} &  & \textbf{+1.54} & -0.00 & +0.06 &\textbf{+0.21} & -0.32 & -0.08 & +0.21 & +0.03 & +0.09 & -0.52 & -0.49 & -0.42 & -0.29 \\
\cmidrule(lr){2-17}
 & \multirow{2}{*}{RB-19} & \textit{Spoof} & \multirow{2}{*}{0.302} & +0.17 & -0.00 & -0.00 & -0.00 & +0.01 & +0.02 & +0.05 & +0.00 & +0.00 & \textbf{+0.57} & +0.36 & \textbf{+0.67} & +0.40 \\
 &  & \textit{Both} &  & \textbf{+2.79} & +0.31 & +0.02 & +0.15 & -0.14 & -0.12 & -0.02 & +0.07 & +0.16 & -0.20 & -0.30 & -0.20 & -0.30 \\
\cmidrule(lr){2-17}
 & \multirow{2}{*}{DA-19} & \textit{Spoof} & \multirow{2}{*}{0.442} & -0.11 & +0.01 & +0.00 & +0.00 & +0.01 & +0.00 & -0.02 & +0.02 & +0.04 & +0.29 & +0.23 & +0.25 & +0.21 \\
 &  & \textit{Both} &  & \textbf{+3.23} & +0.01 & -0.02 & +0.02 & +0.05 & +0.33 & \textbf{+0.66} & -0.12 & -0.19 & -0.43 & -0.41 & -0.37 & -0.36 \\
\cmidrule(lr){2-17}
 & \multirow{2}{*}{RB-24} & \textit{Spoof} & \multirow{2}{*}{0.277} & +0.15 & +0.04 & +0.00 & -0.00 & +0.03 & +0.03 & +0.04 & +0.02 & +0.05 & \textbf{+0.33} & +0.08 & \textbf{+0.37} & +0.05 \\
 &  & \textit{Both} &  & \textbf{+3.12} & +0.18 & -0.02 & +0.14 & +0.02 & +0.37 & \textbf{+1.37} & -0.14 & -0.20 & -0.32 & -0.14 & -0.18 & +0.39 \\
\cmidrule(lr){2-17}
 & \multirow{2}{*}{DA-24} & \textit{Spoof} & \multirow{2}{*}{0.481} & -0.12 & +0.03 & +0.02 & +0.00 & -0.02 & -0.09 & -0.15 & +0.03 & +0.05 & -0.15 & -0.15 & -0.17 & -0.16 \\
 &  & \textit{Both} &  & \textbf{+2.46} & +0.25 & -0.03 & +0.09 & +0.86 & \textbf{+1.84} & \textbf{+2.53} & -0.10 & -0.17 & +0.81 & +0.84 & \textbf{+1.43} & \textbf{+1.30} \\
\bottomrule
\end{tabular}
}
\caption{Relative degradation $\delta_{m,p}$
(Eq.~\ref{eq:rel_deg}) under controlled interventions.  Columns report individual perturbations
grouped by category: \textit{Spectral} (S1: band-cut 0--2\,kHz, S2:
2--5\,kHz, S3: 5--8\,kHz, S4: downsample 8\,kHz),
\textit{Noise/Amp} (N1: AWGN 20\,dB, N2: 10\,dB, N3: 5\,dB, N4:
peak norm $\mu{=}0.65$, N5: $\mu{=}0.45$), and \textit{Non-Speech} (T1:
zeros leading, T2: zeros trailing, T3: AWGN leading, T4: AWGN
trailing; all 4.0\,s).  Two intervention targets are compared:
\textit{Spoof}, only spoofed utterances perturbed; \textit{Both}, both classes perturbed.  Base DCF is the clean-condition cost at
the globally optimised threshold $\tau^*$. $*$ denotes frozen SSL front-end.}
\label{table:delta-results}
\vspace{-18pt}
\end{table*}
\vspace{-10pt}
\subsection{Evaluation Metrics}
\label{sec:metrics}

Baseline performance is reported as Equal Error Rate (EER) across all
conditions.  For calibrated decision analysis we use the normalised
detection cost function (DCF)~\cite{9143410} with the cost parameters and spoofing prior defined in the ASVspoof 2019 evaluation
protocol~\cite{ASVspoof19}, consistent across all three evaluation
corpora.  The global threshold~$\tau^*$ is obtained by minimising the
DCF on the clean evaluation set and held fixed across all subsequent
analyses. To quantify the impact of each perturbation we measure the relative
degradation in DCF:
 { 
    \setlength{\abovedisplayskip}{7pt}
    \setlength{\belowdisplayskip}{6pt}
\begin{equation}\label{eq:rel_deg}
    \delta_{m,p} =
        \frac{\mathrm{DCF}_{m,p} - \mathrm{DCF}_{m,\mathrm{clean}}}
             {\mathrm{DCF}_{m,\mathrm{clean}}},
\end{equation}
}%
where $\mathrm{DCF}_{m,p}$ and $\mathrm{DCF}_{m,\mathrm{clean}}$ are
the costs of model~$m$ on perturbed and clean data, respectively, both
evaluated at~$\tau^*$.  Positive values indicate degradation. The
relative formulation makes $\delta_{m,p}$ comparable across
perturbation types within the same model, where the denominator
is constant and the ranking is unaffected by baseline performance.
When comparing across models, we additionally report absolute DCF
values to ensure that large relative values are not artifacts of
near-zero denominators.  We apply this same measure to isolate codec-induced degradation in ASVspoof 2021.

\vspace{-5pt}
\section{Results and Analysis}

This section presents the empirical application of the diagnostic
framework developed in Section~\ref{sec:problem_formulation}. We first establish the generalisation gap to frame our diagnostic question. Next, we test for shortcut dependency through perturbation-based interventions and representational analysis. Finally, we analyse codec and channel-driven degradation, providing contrastive evidence to distinguish true domain shift from shortcut exploitation.


\begin{figure}[!h]
\centering
\includegraphics[width=\linewidth]{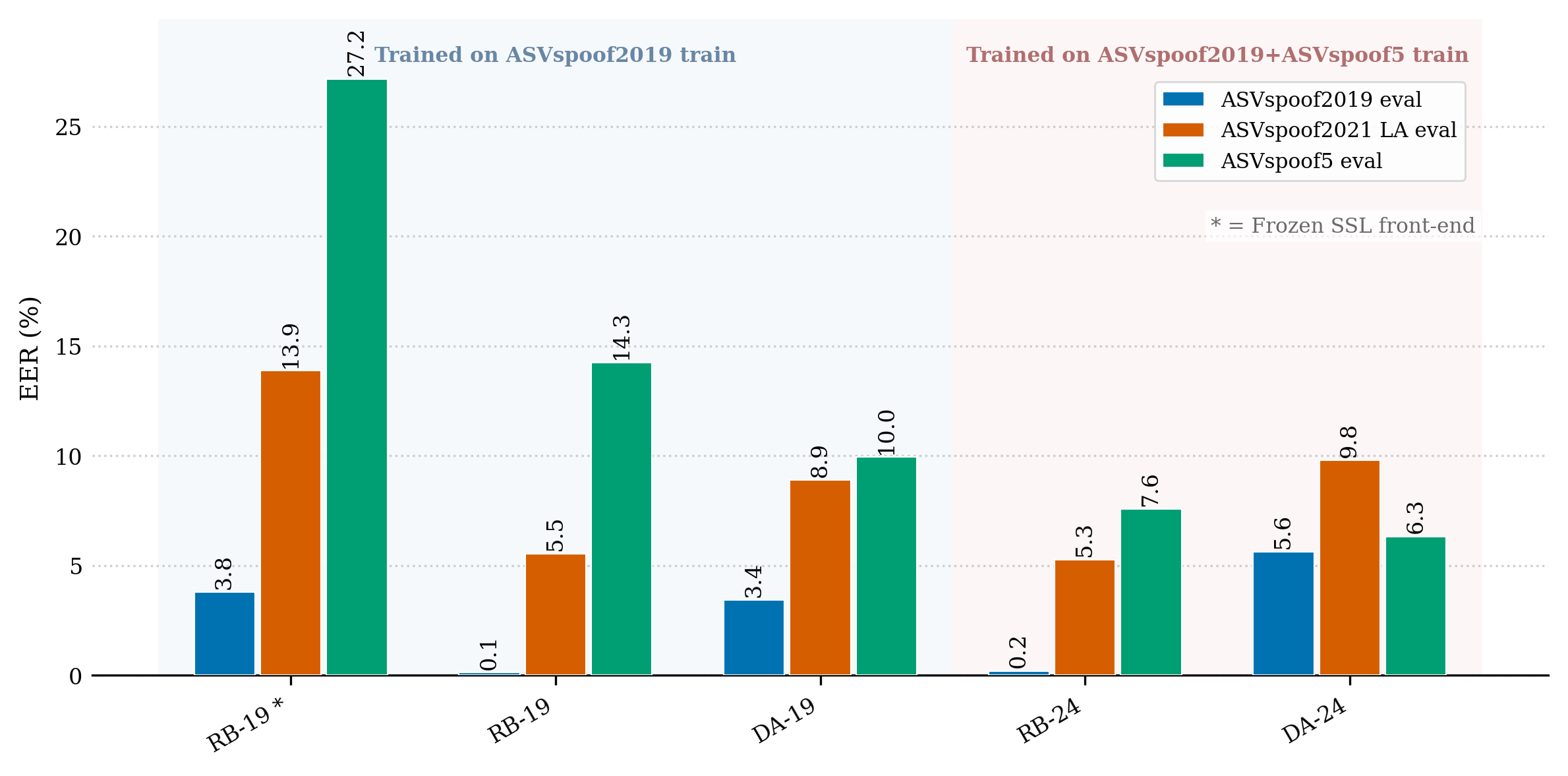}
\vspace{-18pt}
\caption{EER~(\%) across evaluation sets for all models.}
\label{fig:eer_barplot}
 \vspace{-12pt}
\end{figure}

\subsection{Baseline Performance}
\label{sec:baseline}
Figure~\ref{fig:eer_barplot} reports EER across all evaluation conditions for four model configurations, which vary by augmentation strategy (RawBoost [RB] vs.\ custom data augmentation [DA]) and training corpus (ASVspoof 2019 alone [-19] vs.\ combined with ASVspoof 5 [-24]). While all models degrade progressively from ASVspoof 2019 to ASVspoof 5, the magnitudes vary substantially. RB models achieve the lowest in-domain EER but suffer the steepest cross-corpus degradation, whereas DA models trade in-domain accuracy for better cross-corpus robustness. Expanding the training corpus yields expected evaluation gains for both approaches (RB-24, DA-24). Ultimately, however, the choice of augmentation strategy (DA vs.\ RB) dictates generalisation far more than simply adding data. This implies that severe cross-dataset degradation stems primarily from exploiting dataset-specific shortcuts rather than pure acoustic mismatch.

\subsection{Intervention-Based Sensitivity Analysis}
\label{sec:ssi_results}

The generalisation gap in Section~\ref{sec:confounded_shortcuts} stems from either acoustic domain mismatch ($C_i$) or corpus-specific artefacts ($C_d$). Since Section~\ref{sec:corpus_analysis} identified non-speech as the primary condition~(i) shortcut, Table~\ref{table:delta-results} tests condition~(ii) via sensitivity profiling.

 \begin{figure*}
    \centering
    \includegraphics[width=0.9\linewidth]{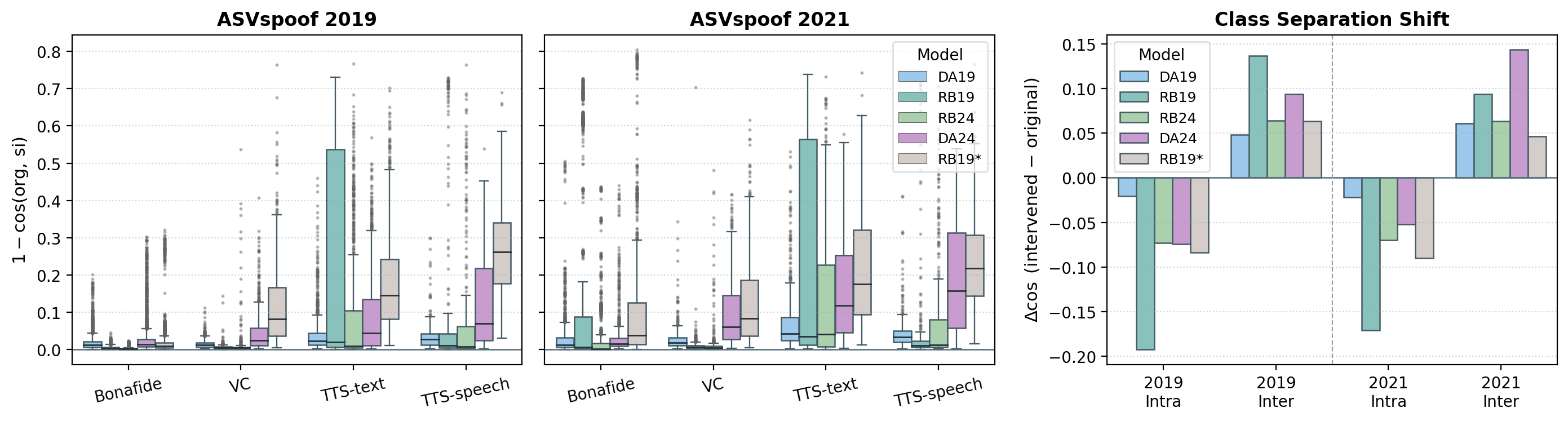}
    \vspace{-7pt}

    \caption{Embedding analysis under non-speech intervention (2,000 balanced utterances). Left/centre: original intra-class cosine distance by input modality synthesis category. Right: intra- and inter-class $\Delta$ cosine similarity (intervention $-$ original). A negative inter-class $\Delta$ indicates opposing classes move closer together.}
     \label{fig:cossim}
     \vspace{-13pt}
 \end{figure*}
\vspace{-12pt}
\paragraph*{\textbf{Low-frequency masking.}}
The 0--2\,kHz band-cut (S1) yields a massive relative degradation ($\delta$) under the \textit{Both} target across all models, including DA. Rather than exposing a dataset shortcut ($C_d$), this universal drop indicates the masking of legitimate generative traces ($Z$). Removing this critical low-frequency content strips away intrinsic speech cues, artificially collapsing the boundary between bonafide and spoofed utterances in the representation space. In contrast, other spectral interventions exhibit consistent, moderate behaviour.

\vspace{-12pt}
\paragraph*{\textbf{Noise sensitivity in joint-corpus models.}}
Models trained on the joint ASV19+ASV5 corpus (RB-24, DA-24) exhibit 
substantially larger degradation under additive noise interventions, 
most markedly under the \textit{Both} target, a pattern consistent 
across both ASVspoof 2019 and 2021 eval conditions. This is 
consistent with the distributional analysis of 
Section~\ref{sec:corpus_analysis}: adding ASV5 training data 
increases the class separability of speech and non-speech power 
descriptors (Fig.~\ref{fig:scatter_jsd}, rightward arrows), 
suggesting that joint-corpus models develop sensitivity to energy 
conditions that ASV19-only models do not exploit.

\vspace{-12pt}
\paragraph*{\textbf{Non-Speech Interventions.}} The clearest pattern in Table~\ref{table:delta-results} is the stark divergence between RB and DA models. RB-19 suffers an extreme degradation ($\delta > 60$) under leading non-speech addition. Because this modification is strictly confined to non-speech regions where genuine generative traces ($Z$) are absent, this performance drop unambiguously reflects reliance on the non-speech shortcut. Conversely, DA models, which are trained with non-speech trimming, show near-zero or negative $\delta$, confirming they are fully decoupled from $C_d$. Crucially, even though RB-24 is trained on a larger corpus (without non-speech augmentation), it still yields $\delta = {+}18.83$. This proves that simply scaling the dataset does not resolve the shortcut if the added data preserves the same confound.

\vspace{-12pt}
\paragraph*{\textbf{Cross-dataset attenuation.}}
On ASVspoof 2021 LA, non-speech $\delta_{m,p}$ drops sharply, RB-19 falls 
from $\delta_{m,p} > 60$ to $\delta_{m,p} = {+}0.67$ for the same intervention, consistent with the protocol-level neutralisation of the 
$C_d$--$S$ association documented in 
Section~\ref{sec:corpus_analysis}. Notably, under \textit{Both}, 
$\delta_{m,p}$ turns negative for most models: the \texttt{only\_speech} 
annotation applied by the ASVspoof 2021 authors \cite{liu2023asvspoof2021}, itself an intervention on the bonafide non-speech distribution, partially 
disrupts the training-distribution profile; uniform non-speech addition 
partially restores it, recovering borderline decisions.

\vspace{-12pt}
\paragraph*{\textbf{Representational Leakage}} 
To verify whether this non-speech bias is hardcoded into the learned representation ($\hat{Z}$), Figure~\ref{fig:cossim} visualises the embedding geometry under intervention. The left and centre panels measure the raw displacement caused by leading non-speech fragments, showing that RB embeddings shift drastically from their original state. Notably, for RB-19, the magnitude of this displacement varies substantially across different attack mechanisms, highlighting how deeply the shortcut's influence is entangled with specific synthesis algorithms. More importantly, the right panel illustrates how this shift corrupts class separation: for RB models in ASVspoof 2019, adding non-speech fragments simultaneously scatters embeddings of the same class (negative intra-class $\Delta$) and pulls opposing classes closer together (positive inter-class $\Delta$). This blurring of the decision boundary is the definitive proof for condition~(ii): rather than isolating intrinsic generative features ($Z$), the model actively relies on non-speech structure ($C_d$) to classify. Had the model successfully ignored the shortcut, its embeddings would remain invariant.


\vspace{-12pt}
\paragraph*{\textbf{Codec and Channel Consistency.}}
Codec and transmission conditions in ASVspoof 2021 LA provide 
the contrastive test for $C_i$-driven domain shift. Unlike 
non-speech interventions, where RB and DA models diverge by 
two orders of magnitude, codec-induced degradation is 
moderate and consistent across configurations: GSM produces 
the highest values (up to $\delta_{m,p} = {+}4.06$) yet RB-19 and 
DA-19 show comparable sensitivity ($+0.81$ vs.\ $+0.68$), 
and G722 is effectively transparent for all models. 
Transmission path effects follow the same pattern, with 
\texttt{mad\_tx} producing the highest degradation across 
all configurations without interaction with 
augmentation strategy. This uniformity is the diagnostic 
signature of $C_i$-driven shift: the exogenous factor 
degrades all models equally since it does not operate 
through the $C_d$--$S$ confound, validating the 
framework's core distinction between domain shift and 
shortcut exploitation. On ASVspoof 5, degradation 
magnitudes are larger due to more aggressive channel 
conditions, yet the relative ordering between models 
remains stable across conditions, confirming the same 
$C_i$-driven pattern at greater domain distance.

\section{Conclusion}

In this paper, we have proposed an intervention-based diagnostic framework that formally distinguishes shortcut learning from legitimate domain shift in spoofing countermeasures. By grounding our analysis in a directed acyclic graph, we derived the necessary conditions to identify when a model abandons the true generative footprint ($Z$) to exploit spurious correlations stemming from idiosyncratic pipeline artifacts ($C_d$), a phenomenon that artificially inflates performance in controlled benchmarks but causes severe degradation in the wild.

Our empirical results, obtained through controlled acoustic perturbations, confirm that non-speech structure acts as the dominant shortcut in standard corpora. Simple interventions in non-speech regions produce extreme degradation ($\delta_{m,p} > 60$) in models trained with standard augmentations (RB), whereas strategies directly targeting the shortcut (such as non-speech trimming in DA) successfully decouple the decision boundary from the $C_d$ artifact. However, this mitigation comes at a cost, as it considerably degrades overall performance compared to the original RawBoost baseline. Crucially, we demonstrate that simply scaling the corpus size by combining datasets is insufficient to resolve shortcut dependency if the training protocol does not explicitly mitigate these spurious correlations.

Looking forward, the separation between generative and idiosyncratic artifacts formalized in this study opens a promising research direction. This framework can guide the introduction of source tracing techniques and the application of minimal information criteria during training. Forcing networks to strictly retain the minimal information necessary to isolate the synthesis footprint, while penalizing the encoding of peripheral approximation to confounding variables, will be a fundamental step toward generating latent representations that are inherently robust and generalizable to unseen attacks.

\section{Acknowledgements}

This work has received funding from MCIN/AEI/10.13039/501100011033 under Grant PID2024-155948OB-C53.

\bibliographystyle{IEEEtran}
\bibliography{Odyssey2026_BibEntries}

\end{document}